\documentclass[useAMS,usenatbib]{mn2e}

\usepackage{graphicx}

\title[Scintillation observations of PSR B0823+26]{Scintillation observations of PSR B0823+26}
\author[M. Daszuta, W. Lewandowski, J. Kijak]{M. Daszuta\thanks{E-mail:
muzzy@astro.ia.uz.zgora.pl (MD)} W.Lewandowski J. Kijak\\
Kepler Institute of Astronomy, University of Zielona Gora, Lubuska 2, PL-65-265 Zielona Gora, Poland\\}
\begin{document}

\date{Accepted ...; Received ... ; in original form }

\pagerange{\pageref{firstpage}--\pageref{lastpage}} \pubyear{}

\maketitle

\label{firstpage}

\begin{abstract}
We present results of the analysis of interstellar scintillation in PSR B0823+26. Observations were conducted at a frequency of 1.7 GHz using the 32-m Toru\'n Centre for Astronomy radio telescope. More than 50 observing sessions, lasting on average 10 h, were conducted between 2003 and 2006. We found interstellar scintillation parameters by means of dynamic spectrum analysis as well as structure function analysis of the flux density variations. We identified two distinctive time-scales, which we believe to be the time-scales of diffractive and refractive scintillation. Our results show that at the given frequency the diffractive time-scale in PSR B0823+26 is $\tau_{diss}$ = $19.3^{+1.7}_{-1.6}$ min, the refractive time-scale is $\tau_{riss} = 144 \pm 23$ min and the decorrelation bandwidth is $B_{iss} = 81 \pm 3$ MHz.
\end{abstract}

\begin{keywords}
pulsars: general - pulsars: individual: PSR B0823+26 -ISM:structure.
\end{keywords}

\section{Introduction}
The theory of the propagation of radio waves through the interstellar medium is constantly evolving. The plasma modulation phenomena affecting radio waves can be described by means of the three-dimensional spatial power spectrum of electron density fluctuations, which is described by a Kolmogorov spectrum \citep{r1990}:
 \[
P_{3n} = C^{2}_{n} q^{-\beta},
 \]

\noindent
where $C^{2}_{n}$ is a measurement of the mean turbulence of the electron density along the line of sight and $q = 2\pi/s$ is the wavenumber associated with the spatial scale of turbulence $s$. This formula holds, assuming that the spatial scale $s$ is between the inner and outer scales of the spatial density fluctuations ($s_{inn}\ll s \ll s_{out}$). The spectral index $\beta$ is believed to be in the range 3-5 \citep[see also ][]{romani}. Kolmogorov theory, which describes the propagation of turbulent energy from large scales to small scales, gives the expected value of $\beta = 11/3$.

The theory of interstellar scintillation (ISS) was developed by \citet{s1968}. The simplest model commonly used for the explanation of scintillation and scattering effects is the \textit{thin screen} model. Although it was shown for the case of several lines of sight that this model does not explain the observed scintillation behaviour \citep[see for example][]{bgr1999}, its simplicity allows for easy understanding of these phenomena as well as an interpretation of the observational results. According to this model, the irregularities of the interstellar medium are located within a thin screen, which is located midway between the pulsar and the observer. Scattering within the screen causes the signal to be spread over a time interval $\tau_{s}$ (called the pulse broadening time) and hence shows a variety of phases over a range $\delta\phi \sim 2\pi f \tau_{s}$. Phase modulation of the signal produces interference patterns at the observer’s plane.

Depending on the size of the wavefront perturbations, one can distinguish two types of interstellar scintillations, the so-called \textit{weak} and \textit{strong} scintillation regimes. The strong scintillations can be divided into two different branches: \textit{diffractive} interstellar scintillations (DISS), arising from small-scale fluctuations \citep[ $10^{6}-10^{8}$ m:][]{cwb1985} and \textit{refractive} scintillations (RISS), which form due to large-scale irregularities \citep[$10^{10}-10^{13}$ m:][]{s1982}. 

\begin{table*}
 \hspace{-400pt}
  \begin{scriptsize}
  \caption{PSR~B0823+26 scintillation parameters at multiple frequencies. We present two values of $V_{iss}$: the one originally published and the recalculated one (marked by an asterisk) obtained using the proper values of the pulsar distance and the scintillation velocity constant $A_v$. Values without uncertainties quoted were presented in such a way by the original authors.}\label{compare}
  \begin{tabular}{@{}ccccccccccc@{}}
  \hline
  Freq.    & $B_{diss}$    &   $\tau_{diss}$    &     $\tau_{riss}$                &    $V_{iss}$    &   $V_{iss}^*$ &      u     &   $\theta_{d}$    &  $\theta_{r}$  &   $log C_{n}^{2}$   &     $\alpha$     \\
  (MHz)    &    (MHz)      &      (min)         &           (min)                  &   (km s$^{-1}$)   &  (km s$^{-1}$)  &            &       (mas)         &      (mas)       &     (m$^{-20/3}$)    &                  \\
 \hline
   74$^{a}$&       -       &         -          &$12^{\mbox{\scriptsize d}}\pm6$&     365         &       -       &     -      &        -          &      -         &          -          &        -         \\
  327$^{b}$&$0.293\pm0.041$&    $2.1\pm0.31$    &           5254                   &       -         & $277\pm16$&$47.8\pm0.8$&   $1.01\pm0.02$   &  $0.01\pm0.03$ &   $-3.24\pm0.03$   &        -         \\
  408$^{c}$&     0.26      &          3.3       &           3456                   &       140       &      149      &     34     &          -        &       -        &           -         &        -         \\
  430$^{d}$&$0.39\pm0.07$  &           -        &             -                    &       -         &       -       &     -      &          -        &       -        &       -3.64         &     4.4          \\
 1000$^{e}$&      33       &       11           &             -                    &   $241\pm72$    &  $206\pm61$   &      -     &          -        &       -        &         -           &        -         \\
 1540$^{f}$&  $82\pm5$     &    $13.6\pm8$      &            576                   &       -         &  $187\pm9$    &      6     & $0.070\pm0.002$   &$0.004\pm0.005$ &        -2.8         &     3.9          \\
 1700$^{g}$&  $81\pm3$     &$19.3^{+1.7}_{-1.6}$&         $144 \pm 23$             &       -         &   $108\pm21$  &     2.7    & $0.065\pm0.002$   &     0.005      &        -2.67        &  $3.94\pm0.36$   \\
 4.75$^{h}$&     -         &     $9\pm1$        &             -                    &       -         &       -       &      -     &          -        &       -        &         -           &        -         \\
10.55$^{h}$&     -         &      $7.5\pm2$     &             -                    &      -          &       -       &      -     &          -        &       -        &         -           &        -         \\
\hline
\multicolumn{11}{l}{$^{a}$\citet{grc1993}, $^{b}$ \citet{bgr1999}, $^{c}$ \citet{sw1985}, $^{d}$ \citet{cwb1985}, $^{e}$ \citet{g1995},}\\
\multicolumn{11}{l}{$^{f}$ \citet{w2005}, $^{g}$ our results, $^{h}$ \citet{m1996}}\\
\end{tabular}
 \end{scriptsize}
\end{table*}

\vspace{0pt}
\begin{figure*}
\includegraphics[trim = 0pt 40pt 0pt -25pt, width=450pt,height=570pt, angle=270]{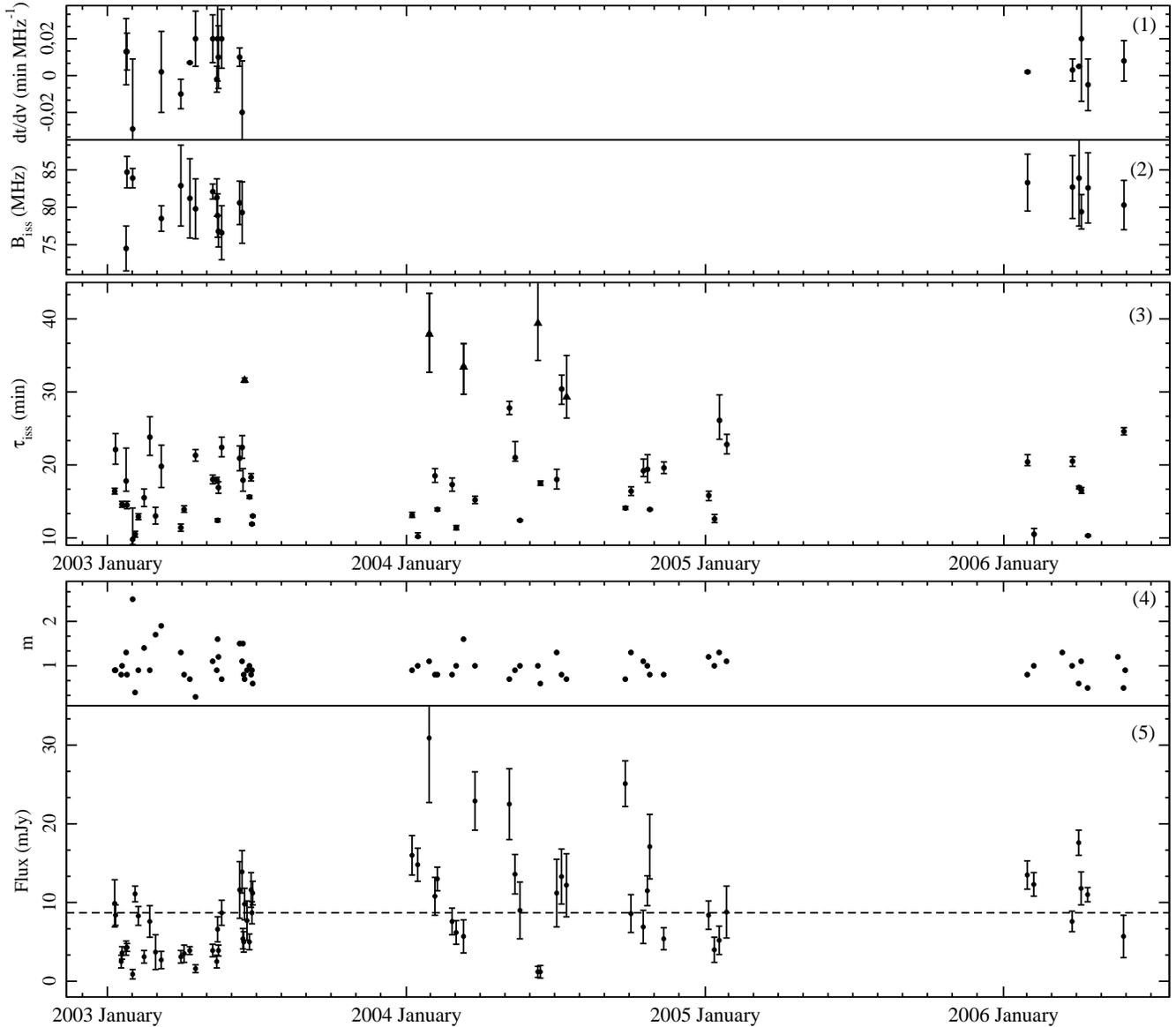}
\vspace{0pt}
\caption{Plots of the time series of scintillation parameters and the PSR~B0823+26 flux density. The drift rate ($dt/d\nu$, panel (1)) and the decorrelation bandwidth ($B_{iss}$, panel (2)) are derived from the dynamic spectrum analysis, while the scintillation time-scale ($\tau_{diss}$, panel (3)) is derived from the structure function analysis. Triangles in the diffractive time-scale plot denote those measurements we do not consider reliable, due to confusion with the refractive time-scale. The last two plots show the  modulation index ($m$, panel (4)) and session-average flux density ($F$, panel(5)). The uncertainties in the measurements indicate $\pm 2 \sigma$ and in the case of the flux density they were either calculated using equation~(\ref{kaspi}) or using the time series RMS (whichever value was greater).\label{flux}}
\end{figure*}
\begin{figure}
\hspace{-20pt}
\vspace{10pt}
\begin{tabular}{c}
\resizebox{\hsize}{!}{\includegraphics[width=160pt,height=247pt, angle=270]{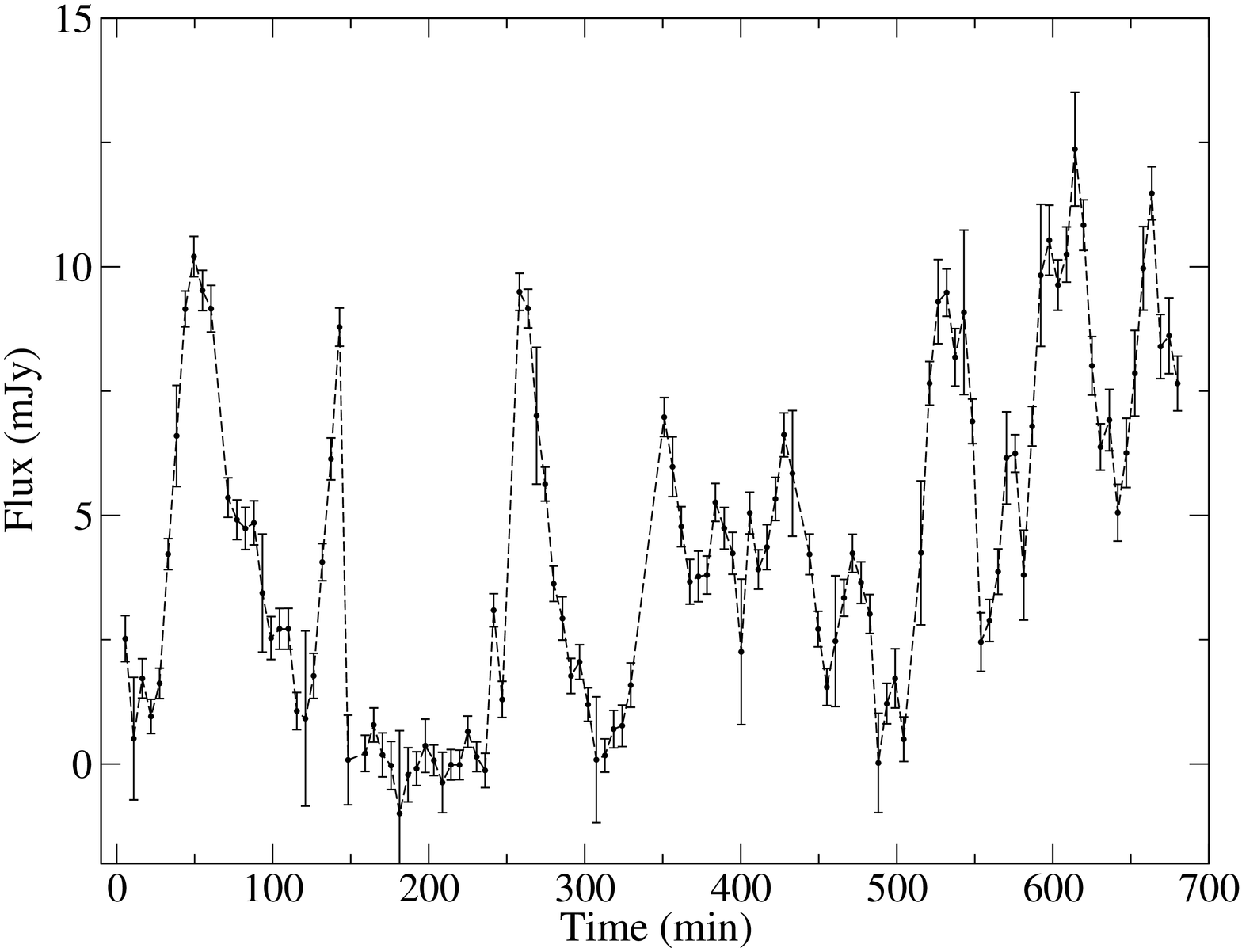}}\\
\resizebox{\hsize}{!}{\includegraphics[width=160pt,height=247pt, angle=270]{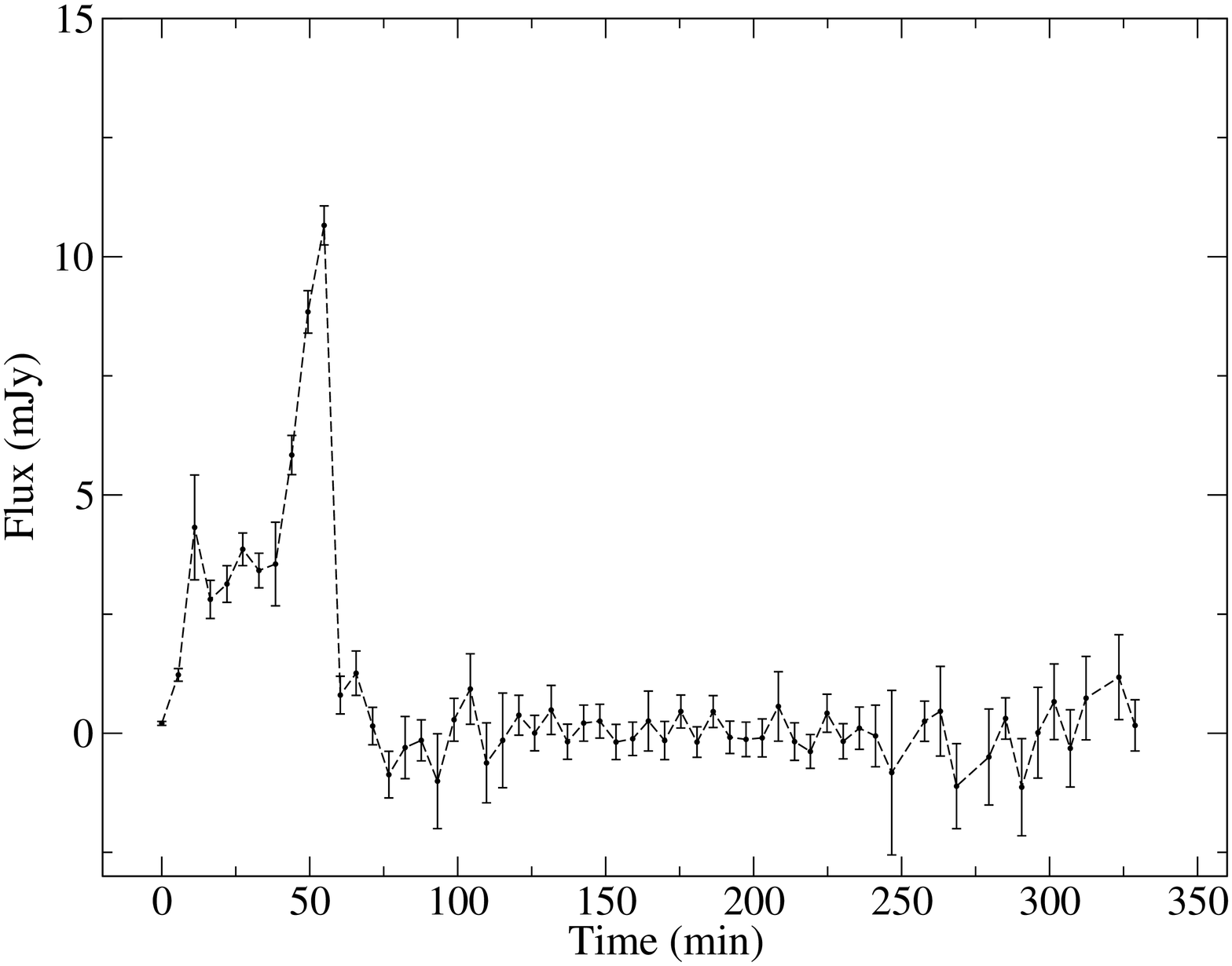}}\\
\resizebox{\hsize}{!}{\includegraphics[width=160pt,height=247pt, angle=270]{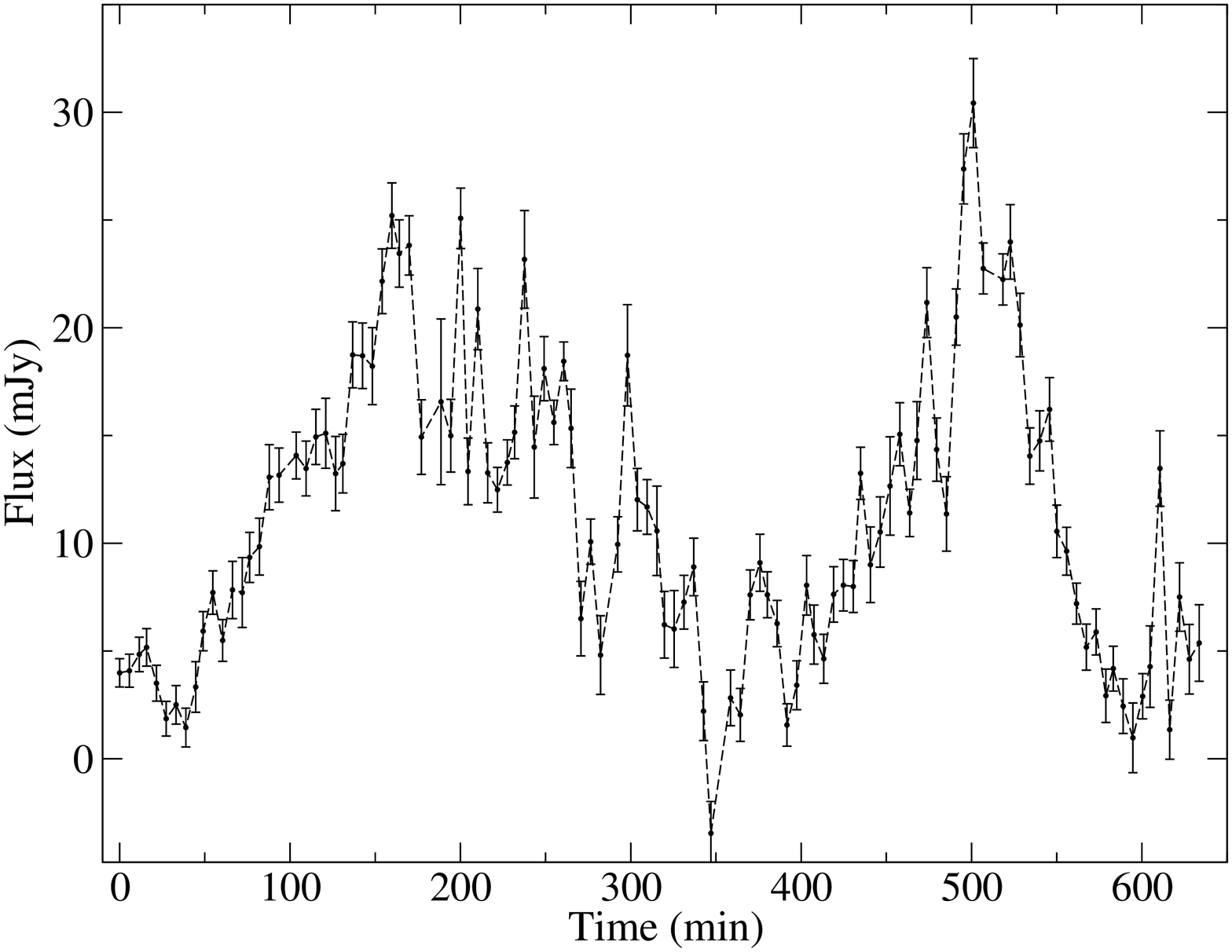}}\\
\end{tabular}
\vspace{-10pt}
\caption{Flux density variations of PSR~B0823+26 during selected individual observing sessions. The first plot shows a light curve with characteristic flat-bottomed minimum with a duration of 100 min (2003 January 23). The middle plot shows another session where the pulsar flux density increases and then suddenly vanishes completely for the reminder of the session (2003 January 31). The bottom plot shows an example of a session with a long ($\sim 200$~min) time-scale clearly visible (2003 June 27). Errorbars for the single-integration flux measurements include instrumental effects.}\label{lens_long}
\end{figure}

The scintillations can be characterized by three basic parameters: the scintillation time-scale $\tau_{\mbox{\scriptsize iss}}$, the decorrelation bandwidth $B_{iss}$ and the modulation index $m$, which is the ratio of root-mean-square (RMS) deviation of the observed flux densities $F$ to the mean value of the time series \citep{lk2005}:
\begin{equation}\label{modulation_index}
   m = \frac{\sqrt{\langle (F-\langle F\rangle)^{2} \rangle}}{\langle F\rangle}.
\end{equation}

The decorrelation bandwidth is related to the scatter by the relation: $2\pi\tau_{\mbox{\scriptsize iss}} B_{iss} = C_1$, where $C_1$ is often assumed to be close to unity, although it may vary for different geometries and/or models \citep{lambert99}.

Another useful parameter, the scintillation velocity $V_{\mbox{\scriptsize iss}}$ is a value describing the velocity at which the observer passes through the scintillation interference patterns. The scintillation velocity combines the transverse pulsar velocity, the velocity of the Earth's transverse motion and the intrinsic screen velocity. This value can be estimated using other scintillation parameters as:
\begin{equation}\label{vis}
  V_{iss} = A_{\mbox{\normalsize v}} \frac{\sqrt{d_{\mbox{\scriptsize kpc}} B_{\mbox{\scriptsize iss,MHz}} X}}{f_{\mbox{\scriptsize GHz}} \tau_{\mbox{\scriptsize iss,sec}}},
\end{equation}

\noindent
where constant $A_{\mbox{\normalsize v}}$ depends on geometry. In our calculations, following \citet{grl1994}, we used the value $A_{V} = 3.85\times10^{4}$ $\mbox{\normalsize km s}^{-1}$. Here, $d_{\mbox{\scriptsize kpc}}$ is the pulsar distance and $X$ is the ratio of the screen-observer distance to the screen-pulsar distance. For a thin screen located midway between pulsar and observer, one obtains X = 1.

In dynamic spectrum analysis, one can often observe the sloped patterns \citep{hwg1985,bgr1999} that arise from refraction. These sloped fringes can be characterized by the frequency drift rate $dt/d\nu$, which can be related to the refractive scattering angle $\theta_{r}$ by the following expression:
\begin{equation}
 \frac{dt}{d\nu} = \frac{\theta_{r}d_{\mbox{\scriptsize kpc}}}{V_{\mbox{\scriptsize iss}}f_{\mbox{\scriptsize GHz}}}.
\end{equation}

In this article, we present the results of our analysis of ISS observations of PSR~B0823+26, at a frequency of 1.7 GHz. Our observations were longer by a factor of four than any other observing project conducted for this pulsar in the past, since we had over 600 h of integration time for this pulsar. 

\section[]{Observations and data analysis}\label{datalys}
The observations were conducted at a frequency of 1.7 GHz using the 32-m Toru\'n Centre for Astronomy radio telescope. The total intensity signal was obtained using the Penn State Pulsar Machine II (PSPM II:\citep{k1999}), which is basically a 64-channel filter-bank spectrometer with a total bandwidth of 192 MHz. Calibration of the signal was performed by injecting the noise-diode signal to our receiver, which was carried out synchronously with the pulsar period. At the start and end of each observing session, the noise diode signal itself was calibrated by means of total intensity observations of a well-known radio source 3C393.1.

The PSPM II output (in the observing mode we used) was basically 64 time-integrated pulse profiles corresponding to the backend spectral channels. This allowed us to perform both dynamic spectrum analysis and – after the dedispersion process and an offline integration of the whole bandwidth – total intensity (pulsar average flux density) measurements.

Observations were conducted between 2003 and 2006. We gathered data during 70 observing sessions, consisting of 5 min individual integrations. More than 50 of those sessions lasted for 10 h or more; there were also shorter sessions (lasting less than 6 h) and the longest sessions lasted for more than 12 h. The total integration time of all our observations analysed for the purpose of this article was $\sim600$ h.

The pulsar we observed was PSR~B0823+26. This is relatively well-studied source, but its scintillation properties have never before investigated in such a long-term extensive observing project. It is a typical, 530-ms period pulsar, which is relatively close at a distance of 0.38 kpc \citep{g1986}, with a dispersion measure of $DM = 19.4$~pc~cm$^{-3}$. The scintillation parameters of the PSR~B0823+26 have been measured previously by many authors at various frequencies, ranging from 74 MHz to 10.55 GHz (see Table~\ref{compare} for a full list of references).

\subsection{Flux density measurements}
The bottom panel of Fig.~\ref{flux} presents the flux density averaged over the individual observing sessions versus observing epoch. As one can see, single-session average values range from $\sim$ 1 mJy to over 30 mJy. To estimate the errors of the average flux measurements we used two methods. First we used an equation that considers the physical scintillation effects \citep{ks1992}:
\begin{equation}
\label{kaspi}
 \frac{\delta F}{F}\simeq\frac{m_{riss}}{\sqrt{T_{obs}/\tau_{riss}}},
\end{equation}
\noindent
where $T_{obs}$ is the total observation time, $\tau_{riss}$ is refractive timescale (we used $\tau_{riss}=144$ min, see the following sections) and $m_{riss}=0.6$ is the expected RISS modulation index (based upon \citealt{lk2005}: $m_{riss} = (B_{iss}/f)^{1/6}$). For the second method, we estimated RMS for each individual session (with the receiver/calibration noise factor included), which turned out to be up to 30 per cent of the average flux value. As a final uncertainty estimate for a given observing session we used the greater of the two values; these are shown as error bars in the bottom panel of Fig.~\ref{flux}. We also calculated the total average flux value based on all available individual integrations, which yielded a value of $\langle F\rangle$ = $8.7\pm1.5$ mJy. The flux density is strongly modulated in the individual sessions: panel (4) in Fig.~\ref{flux} shows the value of the modulation index (obtained using the formula given in equation \ref{modulation_index}).

The variation of the flux density during individual sessions usually showed the expected quasi-periodic fluctuations caused by the scintillations; however, in a few of our observing sessions we found a characteristic disappearance of the pulsar signal. The upper panel of Fig.~\ref{lens_long} shows an example of a light curve with a flat-bottomed minimum between 150 and 250 min. A similar suppression in quasar light curves was observed by \cite{cfl1998}. They believe that the extinction of the flux corresponds to inhomogeneities of the interstellar medium in the form of a plasma lens (with a Gaussian distribution of density of free electrons) crossing the line of sight of the quasar (or, in principle, any point source). The disappearance of the source would be accompanied by a caustic increase of the flux just prior to and immediately after the minima, due to the increase in flux caused by the reflection of radiation by the incoming/outgoing plasma lens. A detailed study of such occurrence could allow for estimation of the size and possibly velocity of such plasma lenses; however, we found our data to be insufficient for that purpose. Only in one case were we able to observe the full extent of such an event, lasting for $\sim$2 h (top panel of Fig.~\ref{lens_long}), but this particular observation is strongly affected by diffractive scintillations, which make finding the flux baseline - and hence the whole \citet{cfl1998} model - almost impossible. We also noted a few cases where the pulsar disappeared during the session (with a flux increase just prior to it) and was not detectable for the reminder of the session, i.e. the disappearance was at least 6-7 h long (see the bottom panel in Fig.~\ref{lens_long}). We also detected one case where the pulsar was not visible at the beginning of the session but then suddenly appeared after a few hours (apparently during a single 5 minute integration) with a significant flux density value.

Overall, such occurrences appear not to be very common for PSR~B0823+26 (only four of our 70 sessions); when such an occurrence appears, its duration is from 2 h up to intervals comparable to the length of our observing session, or possibly longer.

\subsection[]{Analysis of dynamic spectra}
The dynamic spectrum is a measure of the intensity of the pulsar at multiple frequencies over time. This is best described via a two-dimensional image of the pulse intensity as a function of time and frequency. The pulsar is a source emitting coherent radiation; diffraction and refraction in the interstellar medium disturb the wavefront, which leads to self-interference, amplifying or suppressing the pulsar signal as a result. This can be seen in the dynamic spectrum as changing bright and dark bands or patches, which form a pattern of diffractive scintillations of a pulsar \citep[see for example ][]{cpl1986}. A single pattern (enhanced region of flux density) in a dynamic spectrum is called a scintle.

Using the dynamic spectrum, one can estimate the decorrelation bandwidth $B_{iss}$, decorrelation (diffractive) time-scale $\tau_{iss}$ and drift slopes $dt/d\nu$. The values of the scintillation parameters are determined by fitting a two-dimensional Gaussian function to the dynamic spectrum auto-correlation function. Using standard given by \citet{c1986}, we define the decorrelation bandwidth $B_{iss}$ as the half-width at half maximum of the intensity auto-correlation function. The scintillation time-scale $\tau_{iss}$ is defined as the half-width at $1/e$ along the time lag axis. The drift rate is a measure of the slope of the line joining the points on the ellipse with the highest correlation at a given frequency offset \citep{grl1994} and can be determined by finding the inclination angle of the fitted two-dimensional Gaussian.
We observed 20 dynamic spectra of PSR~B0823+26. Fig.~\ref{dsf} shows a sample of our dynamic spectra. The $x$-axis represents the frequency (64 spectral channels, each 3 MHz wide, centered at 1.7 GHz), and the $y$-axis is the epoch quantified by the 5-minute length of a single integration.

At the time of our observations the telescope was affected by a wide range of radio interference, both narrow-band persistent features, which basically destroyed all the data in several spectral channels of PSPM II, and short-time wide-bandwidth interference, capable of disturbing an individual integration over the entire observing bandwidth. Most interference cleaning was performed in the offline analysis process.

Since the expected variations in the dynamic spectrum due to the scintillation phenomenon should be rather smooth, then any sharp or abrupt change in the measured flux density (with regard to both time and observing frequency) can be attributed to radio interference. We removed such measurements from our dynamic spectra and replaced them by values interpolated from the adjacent measurements, which were not affected by  radio interference.

The results of our analysis of dynamic spectra are shown in Fig.~\ref{flux} (top two panels). We also calculated the average values of the diffractive time-scale, $\tau_{iss}= 19 \pm 3$ min, the decorrelation bandwidth, $B_{iss} = 81 \pm 3$ MHz, and the drift rate, $dt/d\nu = 0.027 \pm 0.013$~min~MHz$^{-1}$.
\begin{figure*}
\begin{minipage}{175mm}
\centering
\begin{tabular}{cccc}
\includegraphics[trim = 0pt 10pt 0pt -25pt, width=114pt,height=140pt, angle=0]{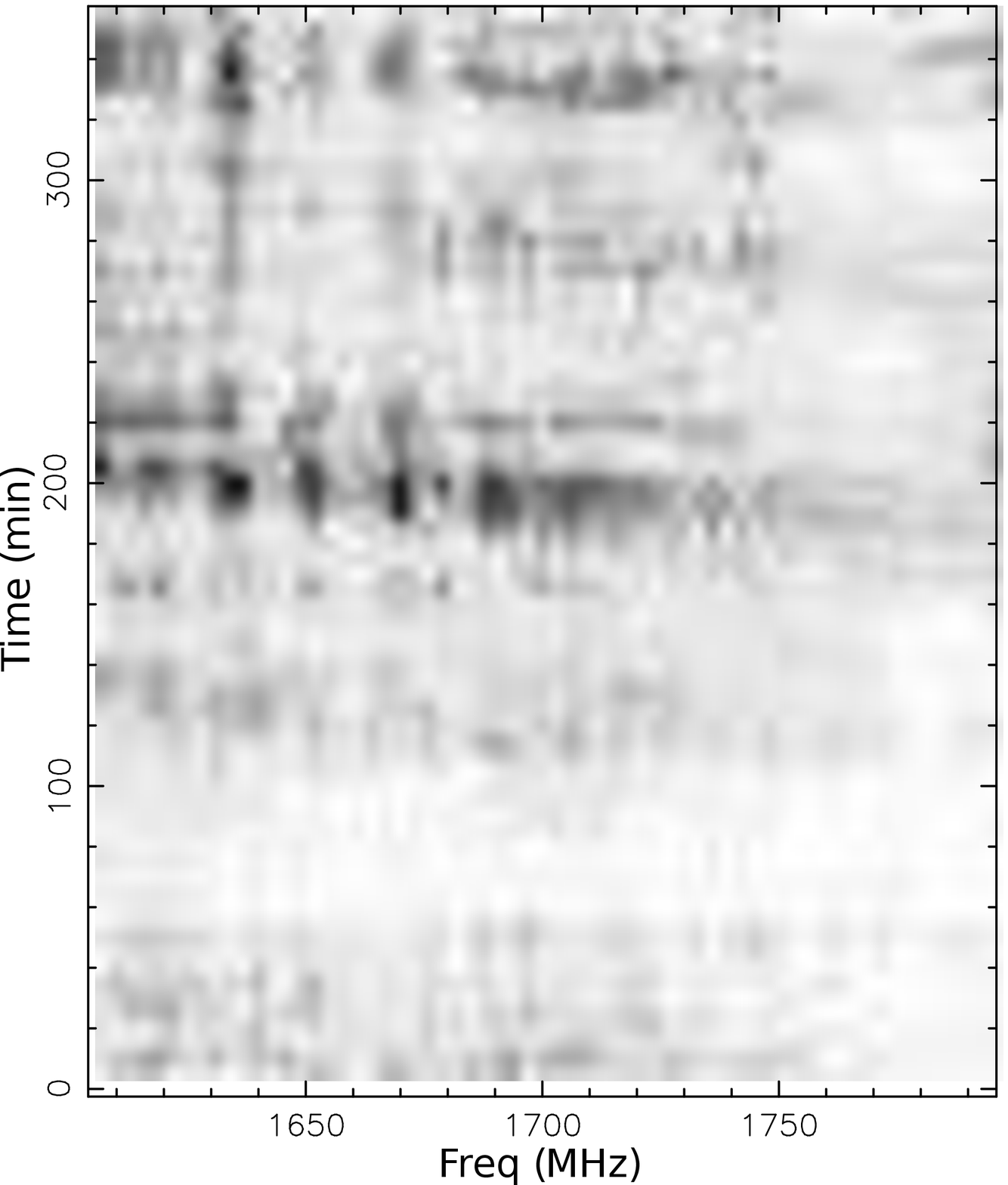} &
\includegraphics[trim = 0pt 10pt 0pt -25pt, width=114pt,height=140pt, angle=0]{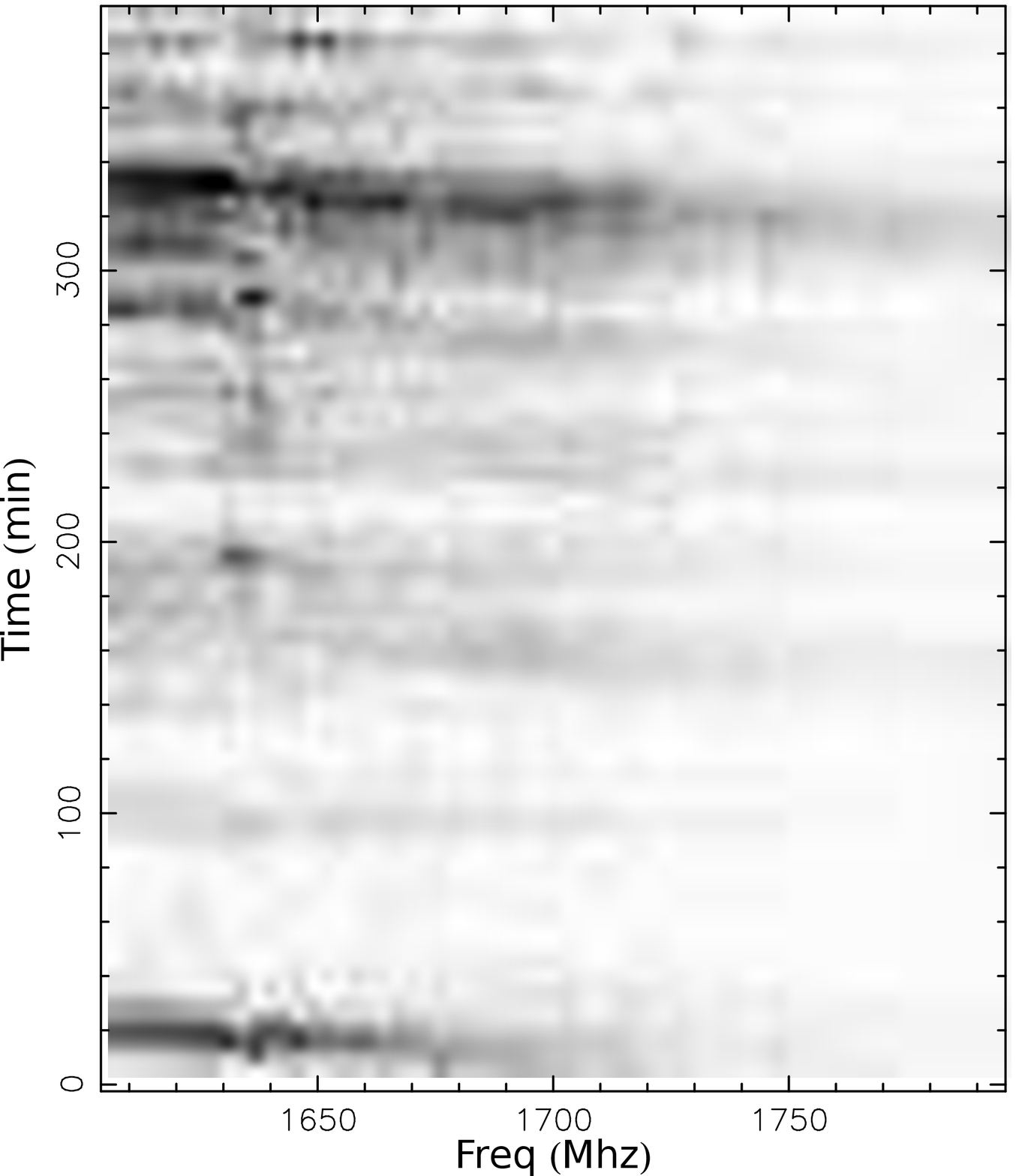} &
\includegraphics[trim = 0pt 10pt 0pt -25pt, width=114pt,height=140pt, angle=0]{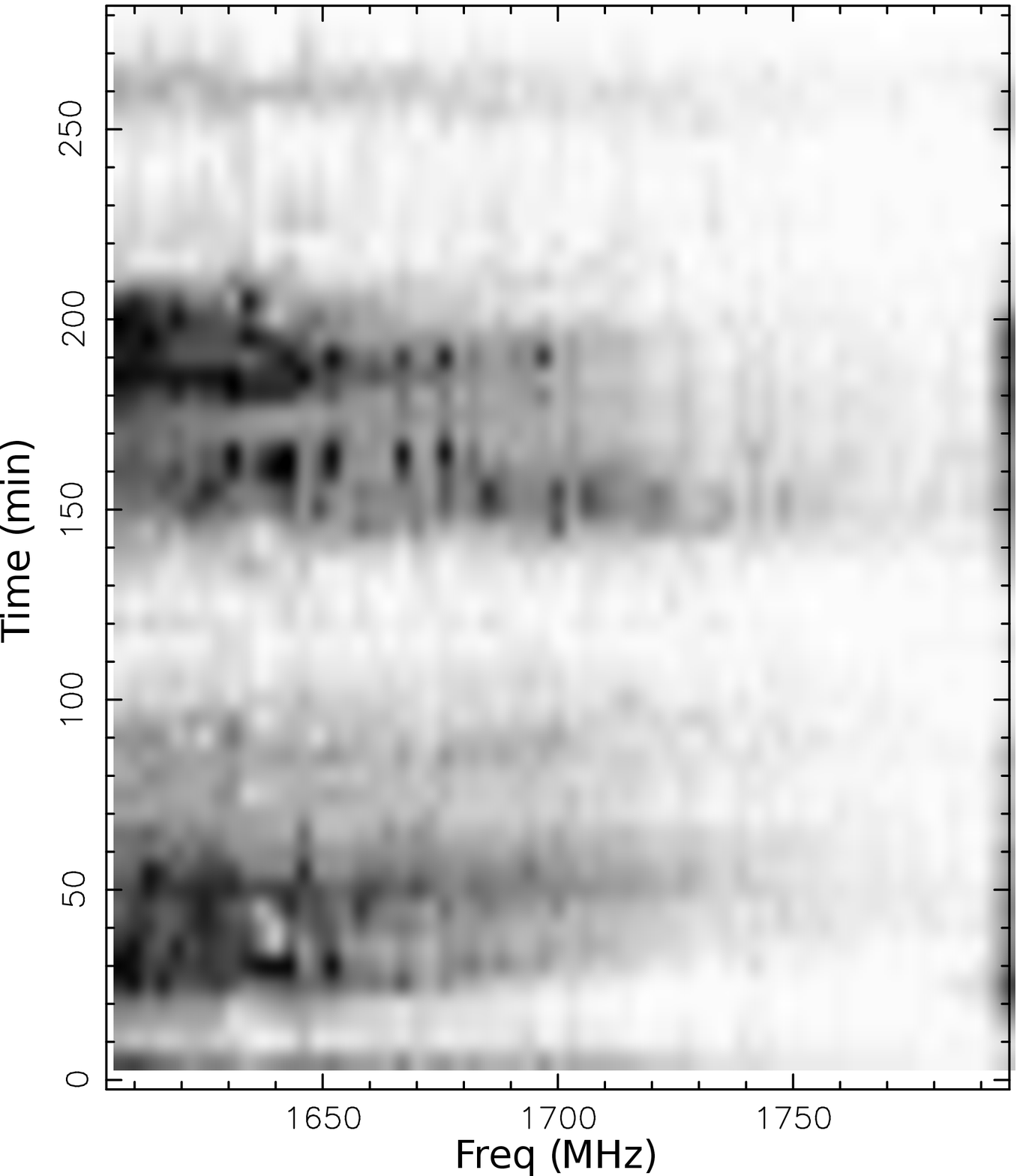} &
\includegraphics[trim = 0pt 10pt 0pt -25pt, width=114pt,height=140pt, angle=0]{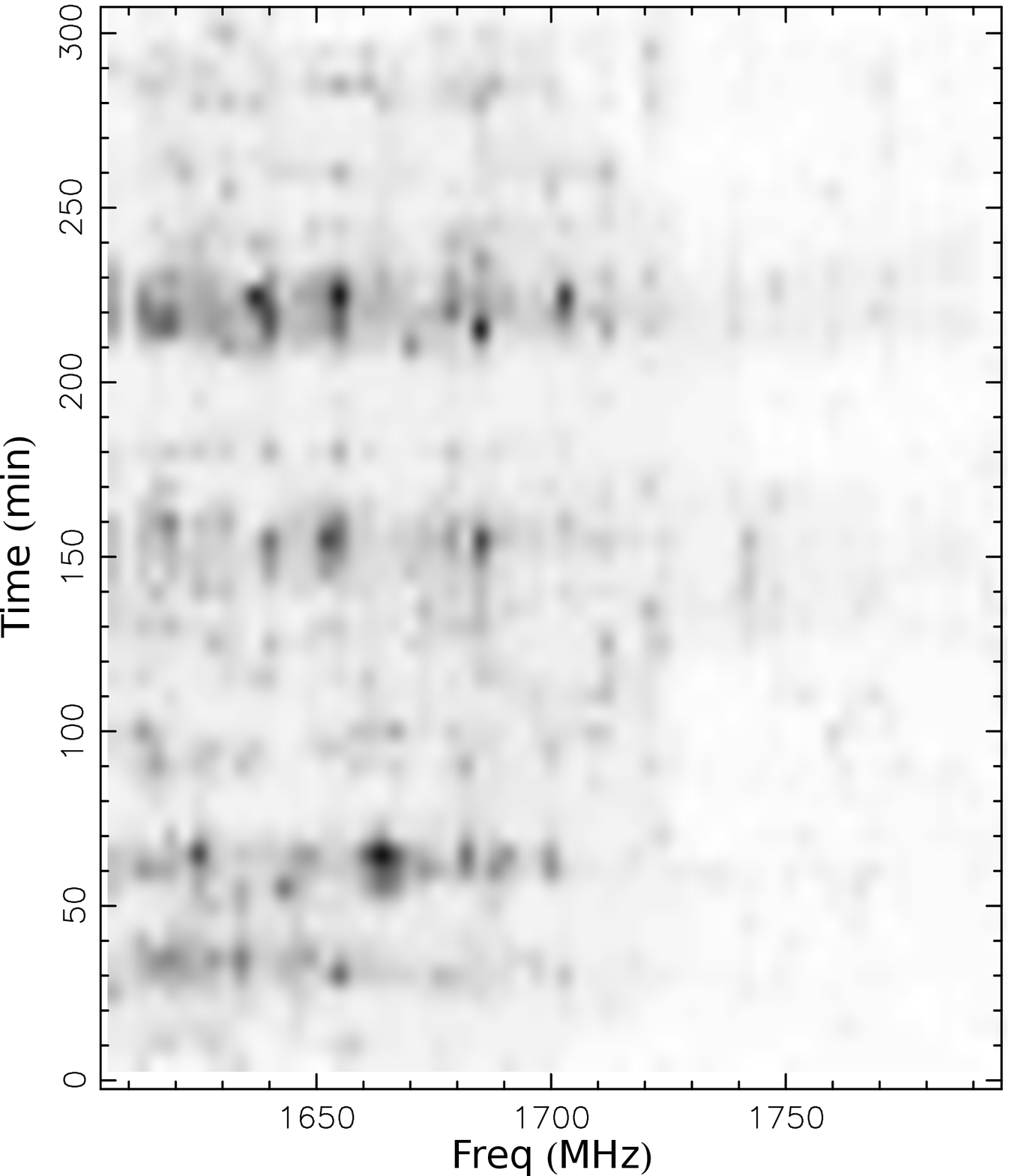} \\
\end{tabular}
\end{minipage}
\caption{Four example dynamic spectra of PSR~B0823+26. The intensity variations of pulsar radiation are presented in the form of gray-scale plots (where a darker shade of gray indicates greater intensity). Starting from the left panel, plots are from 2003 April 18, 2003 May 9, 2003 May 20 and2006 April 14.}
\end{figure*}
\subsection[]{Structure function analysis}
 \label{sfun}
 
\begin{figure}
\vspace{-13pt}
\begin{tabular}{c}
\hspace{-19pt}
\vspace{-10pt}
\includegraphics[width=170pt,height=250pt, angle=270]{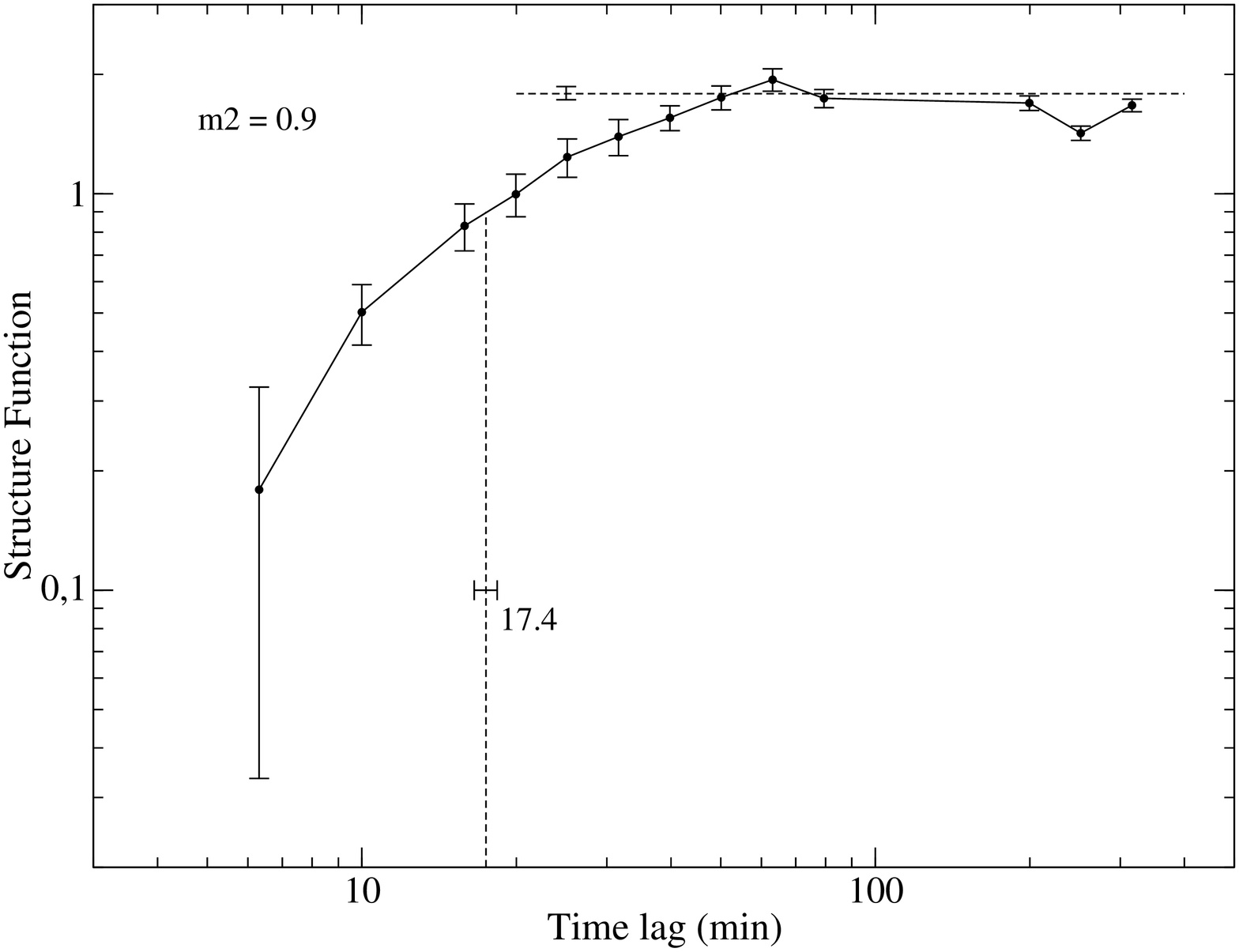}\\
\hspace{-19pt}
\vspace{-6pt}
\includegraphics[width=170pt,height=250pt, angle=270]{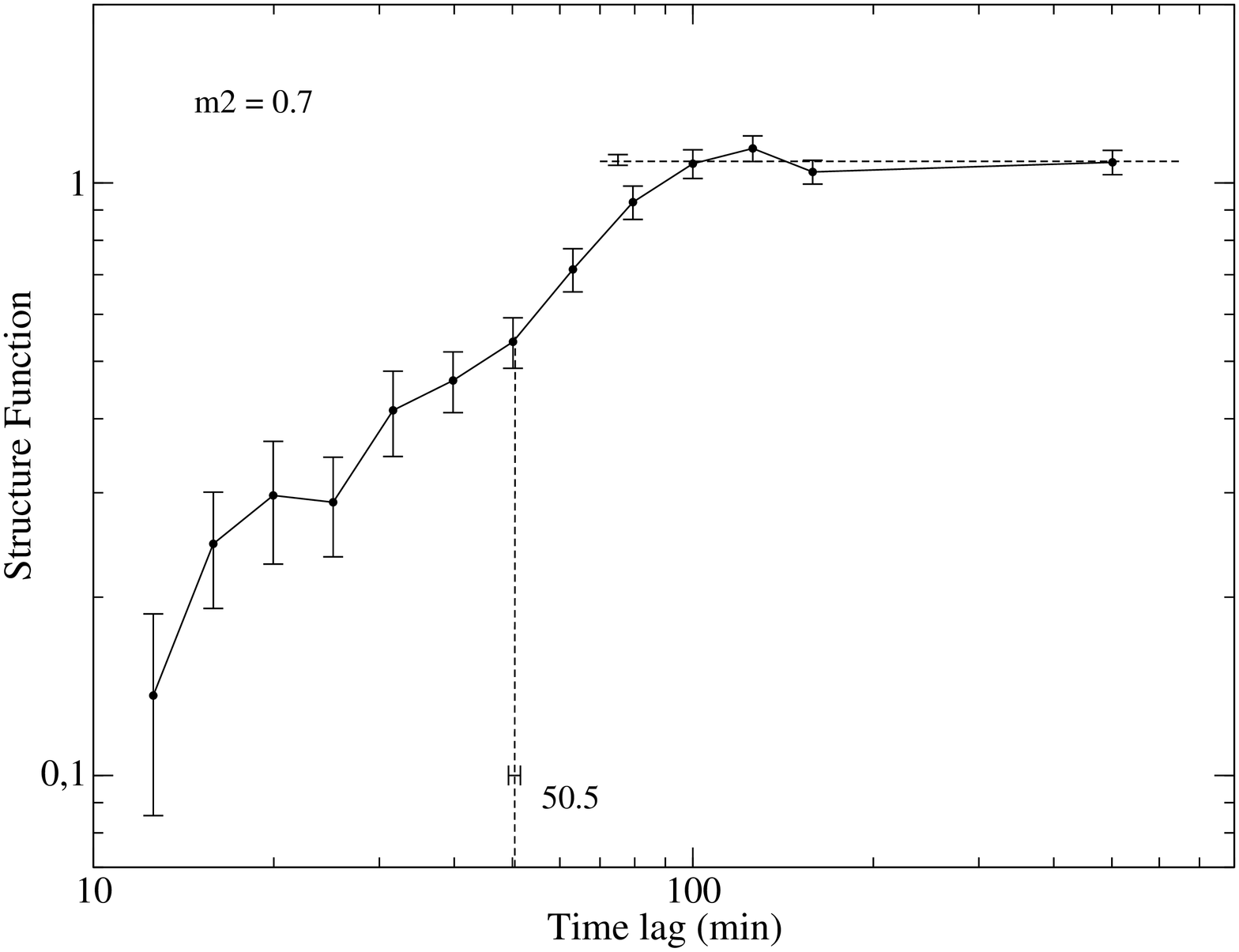}\\
\hspace{-11pt}
\vspace{-5pt}
\includegraphics[width=172pt,height=250pt, angle=270]{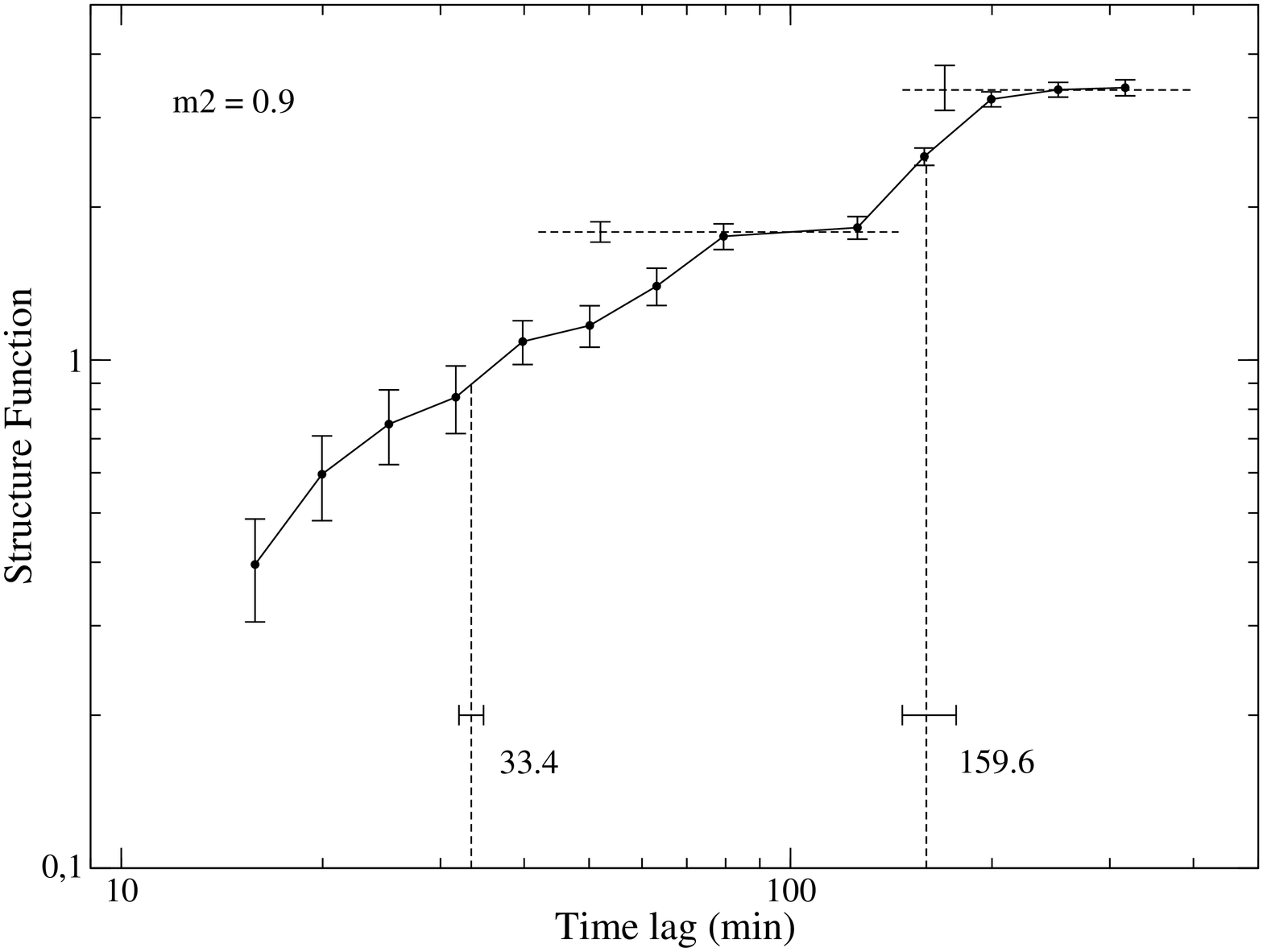}\\
\end{tabular}
\caption{Structure function plot of the flux density time series for two of the individual sessions from our project: 2004 February 26 (top), 2003 June 27 (middle) and 2003 January 18 (bottom). The horizontal dashed line on the plot shows the position of the saturation plateau and the vertical line indicates the value of the scintillation time-scales; m2 denotes the value of the flux density modulation index. Uncertainties in the structure function values were calculated as $\sigma_{D}(i)=\sigma_{\rm WN}\sqrt{8 D(i)/N(i)}$ \citep[where $\sigma_{\rm WN}$ is the white noise: ][]{sch1985}}\label{dsf}
\end{figure}

Structure function analysis is used to study interstellar scintillation and turbulence in random media, as it is an useful tool to estimate the characteristic time-scales of the variability present in the data. We obtained structure function following the method described by \citet{sch1985}. The structure function displays three characteristic regimes.
\begin{enumerate}
 \item The \textit{noise regime} at small lags, dominated by noise and correlated variation of the time series. The time lags are much shorter than the shortest time-scale present in random process.
 \item The \textit{structure regime}, characterized by a monotonic increase, showing a slope in a log-log plot.
 \item The \textit{saturation regime}, where structure function flattens.
\end{enumerate}
The characteristic time-scale of the variations can be found as the time lag at which the value of the structure function is equal to half the saturation value, $D(\tau_{c}) = D_{sat}/2$. Applied to the flux density time series, the structure function analysis can be used to estimate the scintillation time-scale, as well as the modulation index, by means of the formula given by \citep{ks1992}:
\begin{equation}\label{mks}
 m = \sqrt{D_{sat}/2}.
\end{equation} 

The error estimates for the time-scale and the modulation index can be found, assuming that the saturation level uncertainty is $\delta D_{sat}/D_{sat} \approx (2\tau_{c}/T_{obs})^{1/2}$, where $T_{obs}$ is the total duration of the observing session.

We analyzed more than 70 observing sessions using the structure function analysis. Fig.~\ref{dsf} shows typical structure function plots, obtained from individual sessions. Horizontal dashed lines on the plot show the saturation level (with the appropriate error estimates marked), while the vertical lines correspond to the characteristic time-scales found (again, the error bar crossing the dashed line represents the time-scale error estimate). For some of our observing sessions, we were unable to find the scintillation time-scale due to a lack of saturation in the structure function. The results of our measurements of the diffractive time-scale obtained from individual sessions are shown in Fig.~\ref{flux} (panel 3). 

For some of the sessions we were able to detect two plateaus in the structure function plots (see the bottom plot in Fig. \ref{dsf}), which can be attributed to two separate time-scales present in the data. In such cases, the values of $\tau_{\mbox{\scriptsize diss}}$ obtained for the shorter time-scale are shown on the sub-plot (3) and we believe these correspond to the DISS, which (based on the previously published results) at our observing frequency of 1.7 GHz should be of the order of $\sim$ 20 min. The average diffractive time-scale calculated from all the data gathered is $\tau_{\mbox{\scriptsize diss}}$ = $19.3^{+1.7}_{-1.6}$ min. The values marked by triangles on the same subplot are time-scale derived from the second plateau detected in the structure function.

We believe that the second plateau appeared in some of our structure function plots due to the refractive scintillation time-scale. However, since its value is relatively close to the total length of the observing session, the value may be distorted due to insufficient sampling of the structure function at very long time lags. Nevertheless, 
we were able to obtain the value of the time-scale corresponding to the second plateau for five observing sessions, ranging from 108\textendash196 min, with an average of 144 min, which we believe to be a structure function estimate of the refractive time-scale.

One has to note, however, that in some cases the structure function method did not manage to find the second time-scale, despite the fact that it is clearly visible by the naked eye in the flux time series. This is best illustrated by the structure function presented in the bottom panel of Fig.~\ref{dsf}, which is the result of applying the analysis to the time series from the 2003 June 27, shown in the middle panel of Fig.~\ref{lens_long}. One can clearly see that the flux is strongly modulated by - supposedly - refractive scintillations, with a time-scale of $\sim$ 200 min, yet the structure function shows only a single, very broad plateau. Even if we were able to see the refractive time-scale plateau - which should be somewhere behind the right edge of the plot (but was heavily distorted due to insufficient sampling) - its time-scale, clearly visible by eye, would fall within the plateau shown, making the time-scale measurement futile. Hence, for a comparison, we decided to use a simpler method of flux time series auto-correlation function analysis to estimate the value of the refractive scintillation time-scale. Our results, based on a the few best time series where the refractive scintillations were visible, yielded the average value of $\tau_{\mbox{\scriptsize riss}} = 170 \pm 13$~min. This may suggest that the value of the refractive time-scale from structure function analysis (cited in the previous paragraph) may be underestimated. It seems that the refractive time-scale is varying a lot, and since structure function method is unable to detect longer time-scales, the average value obtained from it may be prone to selection effects.

\section[]{Results and discussion}
The theory of the interstellar scintillation predicts that with increasing observing frequency the diffractive time-scale increases, while the refractive time-scale decreases. When observing at the so-called {\it critical frequency}, both time-scales become equal and the scintillations behaviour changes from the strong to weak regime \citep{r1990}. This is best described by the strength of scattering parameter $u$ ($u>1$ for strong scintillations), which can be estimated as \citet{lk2005}
\begin{equation}\label{u2}
 u^{2} = \tau_{\mbox{\scriptsize riss}}/\tau_{\mbox{\scriptsize diss}}.
\end{equation}
Using the value of $\tau_{\mbox{\scriptsize diss}}$ from the structure function analysis, as found from a larger number of observing sessions, we calculated $u = 2.7$.

The strength of the scattering parameter is defined in scintillation theory \citep{r1990} as $u = r_{F}/s_{0}$, where $r_{F}$ is the Fresnel scale and $s_{0}$ is the coherence scale. The latter, in principle, cannot be obtained from observation; one can, however, estimate diffractive and refractive scales. Estimation of the diffractive scale can be made using the value of the scintillation velocity. Using equation\ref{vis} and our results described in the previous section we obtained that $V_{\mbox{\scriptsize iss}} = 108\pm21 $ km s$^{-1}$. The scintillation velocity can be used to estimate the diffractive scale ($s_{d}=V_{\mbox{\scriptsize iss}}\tau_{\mbox{\scriptsize diss}}$), which yielded  $s_{d} = 1.25 \times 10^{8}$ m.

Following \citet{bgr1999}, the diffractive angle can be expressed as  $\theta_{d} = (c/\pi d B_{\mbox{\scriptsize iss}})^{1/2}$, which, using our results, resulted in $\theta_{d} = 0.065$ mas. Using this value, we can estimate the refractive scale ($s_{r} \simeq d\theta_{d}$) as $s_{r} = 3.71 \times 10^{9}$ m. This allows us to estimate the Fresnel scale ($r_F^2=s_d s_r$), which in the case of B0823+26 is $r_{F}= 6.8 \times 10^{8}$ m.

\begin{figure*}
\vspace{-20pt}
\hspace{-17pt}
\begin{minipage}{175mm}
\centering
\begin{tabular}{cc}
\includegraphics[width=161pt,height=230pt, angle=270]{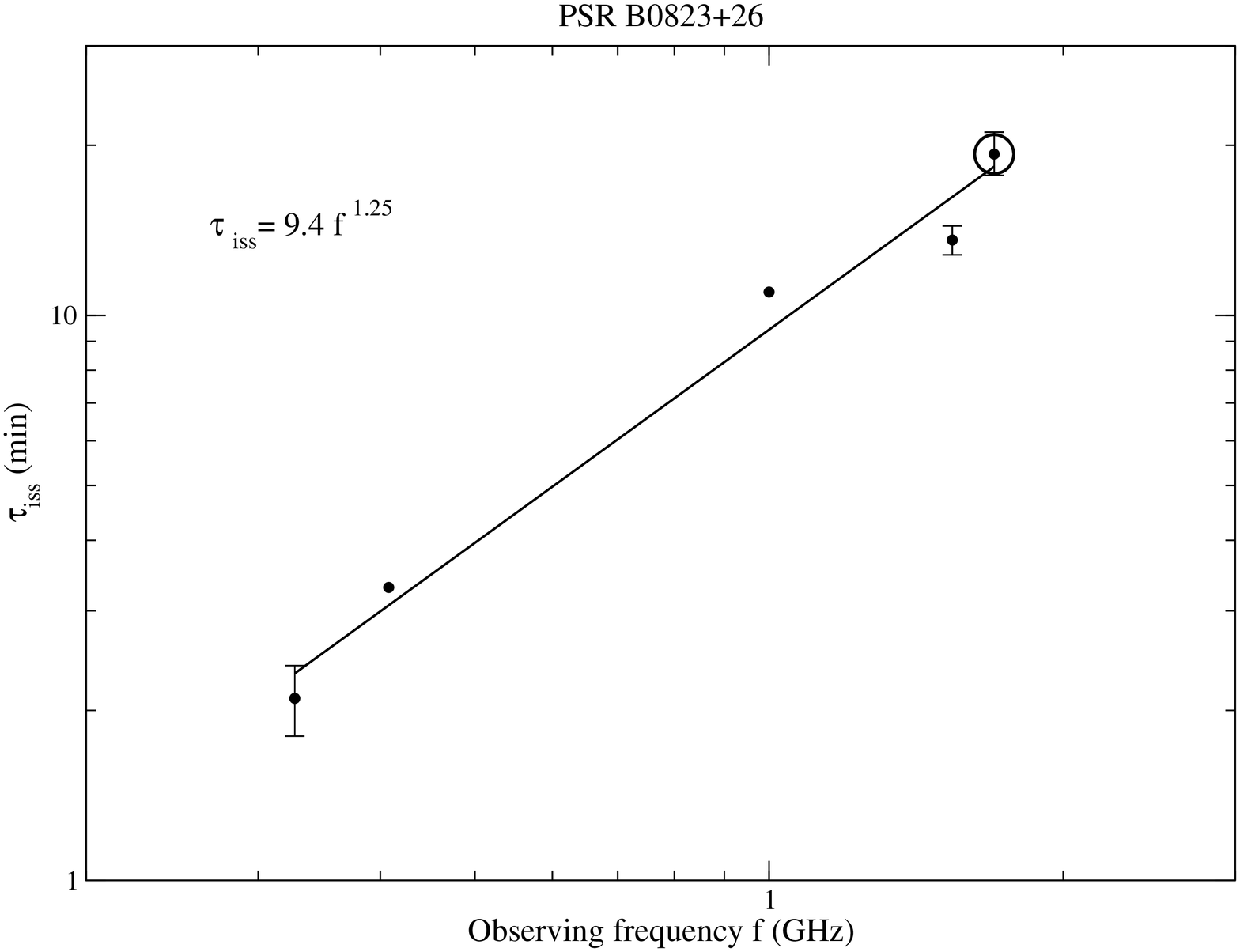}&
\includegraphics[width=161pt,height=230pt, angle=270]{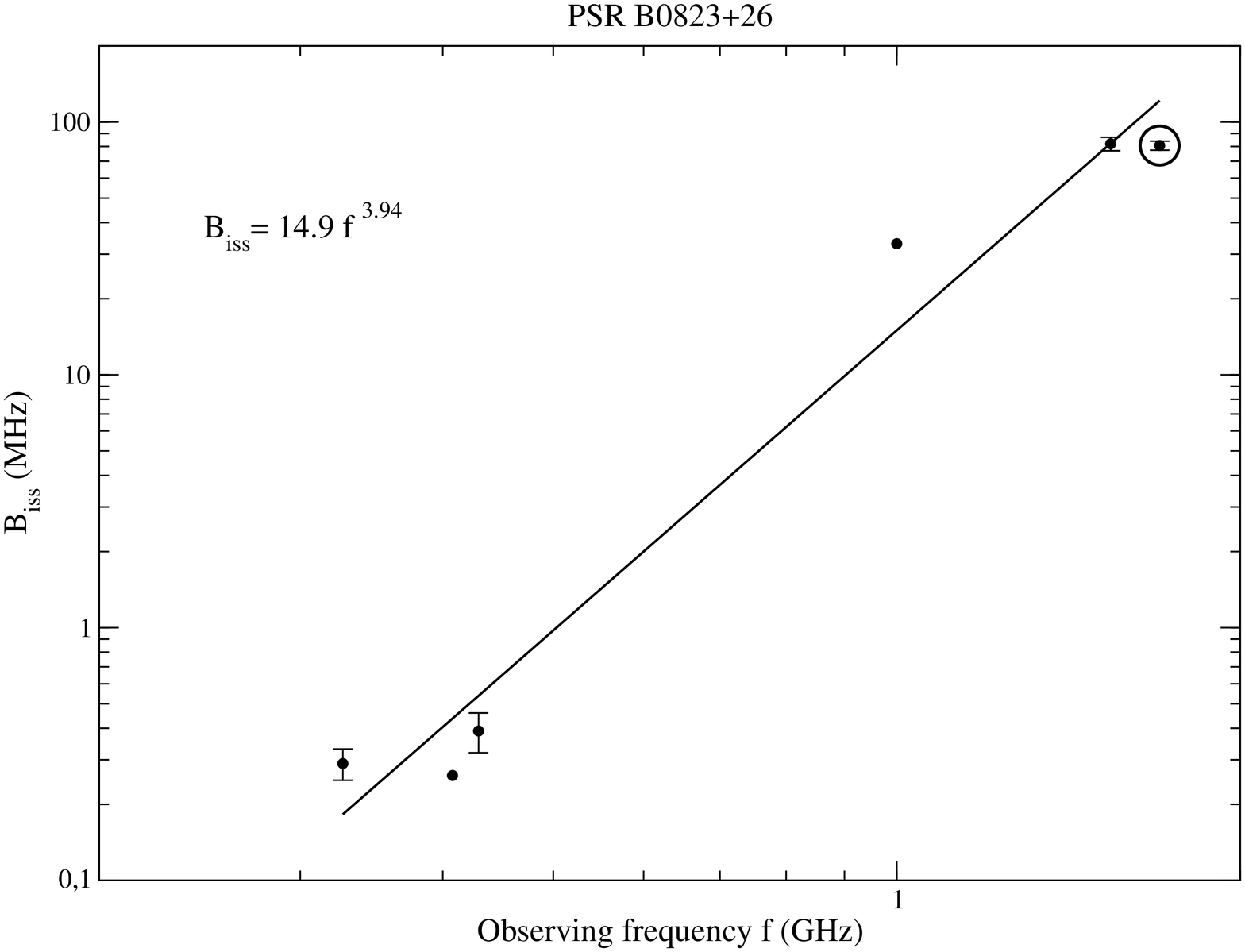}\\
\end{tabular}
\end{minipage}
\caption{Left: a plot of diffractive time-scale versus observing frequency based on data found in the literature and our measurements at 1.7 GHz (circle in the ring). Right: a similar plot for the decorrelation bandwidth. For the references for the lower frequency values, see Table~\ref{compare}. Solid lines represent our fits of the spectral indices (results for which are shown in the upper left part of each plot). Note that some of the values taken from the literature did not have their uncertainties given.}\label{index}
\end{figure*}
\subsection{Spectrum of plasma density fluctuations}\label{spectrum_of_plasma}
Gathering the data from various observing frequencies allows us to study the frequency dependence of the scintillation parameters. The theory of interstellar turbulence predicts that the scintillation parameters depend strictly on the observing frequency \citep{romani}. Table~\ref{powerlow} shows predicted values based various models, as well as our results. Using previously published values (Table~\ref{compare}) and adding our measurements, we estimated the spectral slope for the scintillation time-scale and decorrelation bandwidth (see Fig.~\ref{index}). 

As discussed before, we used value of the DISS time-scale obtained from structure function analysis. A linear fit to the diffractive time-scale versus the observing frequency data yielded $\tau_{\mbox{\scriptsize diss}} \sim f^{1.25 \pm 0.1}$. This is very close to the value expected for Kolmogorov-type turbulence ($\sim$ 1.2). A similar fit for the decorrelation bandwidth versus the frequency dependence yielded $B_{\mbox{\scriptsize diss}} \sim f^{3.94 \pm 0.36}$. One has to note, however, that in both cases a proper uncertainty analysis could not be performed, as for some of the previously published values we could not find the error estimates. Therefore we believe that the decorrelation bandwidth spectral index, although formally inconsistent with the simple model predictions (the value of 4.4 is outside of error estimate), may be marginally considered to be compliant with the Kolmogorov's theory thin-screen model.

The last column of Table~\ref{compare} also includes three values of the decorrelation bandwidth spectral indices published earlier. \citet{sw1985} obtained their value from an independent method: scatter-broadening measurements. \citet{cwb1985} obtained the value from a fit for decorrelation bandwidth (similar to ours, but with fewer data available at the time, only two measurements), and their result is consistent with a Kolmogorov spectrum ($\alpha$ = −4.4). The estimation of \citet{w2005}, as well as our estimate is rather closer to the critical spectrum ($\beta$ = 4) than Kolmogorov one.

Deviations from Kolmogorov’s model predictions are not uncommon: see \citep{boe2013} for a recent summary based on the pulse-broadening observations. In the case of the scintillation studies, \citet{brg1999} showed cases of non-compliance with the simple thin-screen Kolmogorov model in their analysis of 18 objects and more recently \citet{boe2011} provided similar results for PSR~B0329+54, a pulsar that is quite similar to B0823+26 when it comes to the distance and dispersion measure. 

Based on a model for the local interstellar medium, we know that the PSR~B0823+26 is outside the shell of the \textit{Local Bubble} \citep{bgr1998}. Following \citet{cl2002}, the pulsar is also outside the \textit{Loop I}, which is also apparently reflected in its interstellar scattering properties \citep{bg2002}. This could suggest that the interstellar medium density pattern along the PSR~B0823+26 line of sight is indeed completely different from the single thin-screen model, which would explain the possible non-compliance with its predictions.

Using both scintillation time-scales and their frequency dependence (inferred from the fits to the whole observed frequency range data), we were able to estimate that for PSR~B0832+26 the critical frequency is $f_{c}\sim$ 5 GHz and at this frequency one expects a transition from strong to weak scintillation. \citet{m1996} studied the scintillation parameters at 4.75 GHz and 10.55 GHz. Using the structure function analysis - they measured $\tau_{iss}$ = 9 and $\tau_{iss}$ = 7.5 min, respectively. They also suggested that at these observational frequencies the pulsar switched to the weak scintillation regime.

 We have also calculated the level of turbulence (using the formula from \citealt{c1986}) to be $C_{n}^{2} = -2.67$; this value determines the average electron density fluctuation that produce scattering. 

\begin{table}
\begin{center}
 \begin{minipage}{73mm}
\caption{The predicted and observed spectral index of frequency dependencies for the diffractive time-scale and the decorrelation bandwidth.}\label{powerlow}
\end{minipage}
  \begin{tabular}{p{0.1cm}ccccc}
  \hline
 \multicolumn{4}{r}{Predicted spectral index by theory}& Our results\\
                  & ,,steep''& ,,critical''& Kolmogorov&            \\
                  &     4.3  &       4     &     11/3  &            \\
  \hline
 $B_{iss}$        &    +4.7  &      +4     &    +4.4   &$3.94\pm0.36$\\
 $\tau_{diss}$    &    +1.4  &      +1     &    +1.2   &$1.25\pm0.1$\\
  \hline
\end{tabular}
\end{center}
\end{table}
\subsection{Scintillation velocity}
The proper motion and the heliocentric parallax measurements of PSR~B0823+26 provide us with a good estimate of both the pulsar distance and its transverse velocity, which is $V_{pm} = 194\pm41$km/s \citep{las1982}.

Table \ref{compare} also shows the results for the scintillation velocity, which should be close to the pulsar transverse velocity. In the past, various authors used a different distance and the $A_v$ constant (see references) to determine $V_{iss}$. We generalized their measurements (Table~\ref{compare}, sixth column) and recalculated the scintillation velocity using the proper values of the distance $d$ = 0.38~kpc and the constant $A_{v} = 3.85 \times 10^{4}$ (see eq.~\ref{vis}). For our data we estimated $V_{iss}$ = $108\pm21$ km/s. In principle, the values of velocity should be similar, regardless of the observing frequency. However, even after the recalculation one can still note major differences.

One possible explanation may be that our value comes for the grand average of both our DISS time-scale measurements and our decorrelation bandwidth measurements, an average from $\sim 600$ h of integration. However, if we take a values of both $\tau_{diss}$ and $B_{iss}$ from some of our individual sessions, we can get a result similar to the proper motion velocity and close to the value of 220 km/s, which would be in better agreement with the previously published results. Such a discrepancy, especially when it comes from a single - epoch measurements, may arise from the fact that the scintillation velocity is a vector sum of pulsar proper motion, the Earth's orbital motion $V_{E}$ and the bulk flow of the intensity irregularities $V_{irr}$. Depending on the epoch and the screen location, $V_{E}\sim$ may add up to 30 km/s while $V_{irr}$ is usually less than 10 km/s \citep{bgr1999}. 

\section[]{Summary and conclusions} 
In this article we have shown the results of long-term observations of the pulsar PSR~B0823+26 at a frequency of 1.7 GHz. The total integration time of all observations we used was $\sim600$ h, which makes our project the most extensive study of this pulsar's scintillation parameters. Using analysis of dynamic spectra, we found average interstellar scintillation parameters, such as time-scale $\tau_{diss}$ = $19\pm3$ min and decorrelation bandwidth $B_{diss}$ = $81\pm3$ MHz. We also observed characteristic drift patterns with a drift rate of $dt/d\nu = 0.027\pm0.013$~min~MHz$^{-1}$, which indicates the presence of refractive scintillations. Using the values of $\tau_{diss}$ and $B_{diss}$, we estimated the scintillation velocity to be $V_{iss}$ = $108\pm21$~km/s. Our analysis using the structure function method, which we consider more reliable due to a larger sample of data available, yielded a slightly different value of the diffractive time-scale, $\tau_{diss}$ = $19.3^{+1.7}_{-1.6}$ min. 

We were also able to analyze the refractive scintillations characteristics for this pulsar. The dynamic spectrum showed characteristic drifting patterns, which are formed by refraction at large scales. We estimated the time-scale of the variation to be $\tau_{riss} = 144 \pm 23$~min. The measurement of both scintillation time-scales allowed us to estimate the strength of the scattering parameter as $u = 2.7$ and the critical frequency as 5~GHz.

We estimated the derived scintillation parameters such as diffractive scale $s_{d} = 1.25 \times 10^{8}$ m, Fresnel scale $r_{F}= 6.8 \times 10^{8}$ m, diffractive angle $\theta_{d} = 0.065$ mas, refractive scale $s_{r} = 3.71 \times 10^{9}$ m and refractive angle $\theta_{r} = 0.005$ mas.

Addition of our results to those available in the literature allowed us to estimate the frequency dependence of the scintillation parameters in the range 74\textendash1700~MHz. Its  diffractive time-scale varies accordingly to a Kolmogorov-type turbulence spectrum (with either a single thin-screen model or uniform medium distribution), but one may argue that the decorrelation bandwidth does not comply with this. However, since this non-compliance is marginal (see Section~\ref{spectrum_of_plasma}), we believe that PSR~B0823+26 may be considered as a source behaving accordingly to Kolmogorov's theory. 

Summarizing, we analyzed the largest data sample available so far for this pulsar. This scintillation behaviour of this close-by object may be (marginally) explained by a pure single thin-screen model. However, one has to remember that in our analysis we had no choice but to average our values over an entire observational time span, since some of the parameters (like the refractive time-scale) were measurable only for a few individual sessions. Such approach will definitely average out any interstellar medium inhomogeneity effects, if they are present (such inhomogeneities were observed for other low-DM pulsars; see for example \citealt{brisken10}). This may falsify the results of the scintillation parameters and in turn affect the derived spectral indices. We believe that the observed disappearances of the pulsar are the best proof that indeed the interstellar medium along the PSR~B0823+26 line of sight is not homogeneous.

\section*{Acknowledgments}
This paper was supported by the grant DEC-2012/05/B/ST9/03924 of the Polish National Science Centre. We are grateful to all staff at the Toru\'n Centre for Astronomy and especially the TCfA observers group, which conducted the observations. Special thanks to the anonymous reviewer for the comments that helped us to improve the quality of this article greatly.


\label{lastpage}

\end{document}